\documentclass[nolinenumbers]{aa}
\makeatletter
\nolinenumbers
\usepackage{xcolor}
\renewcommand\linenumberfont{\color{white}}
\makeatother

\usepackage{graphicx}
\graphicspath{{./images/}}
\usepackage{amsmath,amssymb,graphicx,bm}
\usepackage{hyperref,float}
\usepackage{natbib}
\nolinenumbers

\graphicspath{{./images/}}

\usepackage{txfonts}
\usepackage{lipsum}
\usepackage{subcaption}         
\usepackage{lscape}             
\usepackage{placeins}           
                                
\begin{document}

   \title{Modified light-cylinder  and centrifugal acceleration in Schwarzschild geometry}

%

  \author{Nikoloz Kurtskhalia\inst{1}\thanks{Email: nkurt22@freeuni.edu.ge}
        \and Nikolai Maltsev\inst{1}\thanks{Email: nmalt22@freeuni.edu.ge}
        \and Zaza N. Osmanov\inst{1,2}\thanks{Email: z.osmanov@freeuni.edu.ge}
}

  \institute{School of Physics, Free University of Tbilisi, 0159 Tbilisi, Georgia}

\institute{$^{1}$School of Physics, Free University of Tbilisi, 0159 Tbilisi, Georgia\\
$^{2}$E. Kharadze Georgian National Astrophysical Observatory, Abastumani 0301, Georgia}

 
  \abstract
   {We examine the motion of an electron constrained to follow a magnetic field line near a primordial sub-stellar mass black hole. Earlier studies treated the problem in flat (Minkowski) spacetime, yielding qualitatively correct results and introducing a \textit{light cylinder} (LC - a hypothetical surface where the linear velocity of rotation exactly equals the speed of light). However, this picture changes significantly when gravity is included. By analyzing the electron’s dynamics in the Schwarzschild metric, we obtain a concept of modified light cylinder (MLC) whose geometry no longer resembles a cylinder. We then determine the maximum energies attainable by the electrons under the limiting effects of inverse Compton scattering, as well as curvature and synchrotron radiation.}


   \maketitle

\section{Introduction}
One of the main challenges of modern astrophysics remains the explanation of the mechanisms that accelerate high-energy particles. There are several key processes, the so-called Fermi mechanism \citep{Fermi1949}  and some of its modifications \cite{Bell1978I, Bell1978II, Catanese1999}. As the analysis shows, the mechanisms listed above share a major drawback: in order for particles to be accelerated to ultrarelativistic energies, they must already be pre-accelerated to certain relativistic energies \cite{rieger}.  Although the Blandford-Znajeck mechanism might be quite efficient, the process is significantly reduced by means of schreening effects \cite{blandford}.  Unlike these processes, the magnetocentrifugal acceleration mechanism does not require pre-acceleration and, at the same time, ensures a very high efficiency of the process.

In the magnetosphere of a black hole, the magnetic field is typically so strong that the plasma is in a frozen-in condition. As a result, gyration of particles  is accompanied by extremely efficient synchrotron radiation, causing the particles to rapidly lose energy and settle into the Landau ground state, continuing to slide along the magnetic field lines. The mechanism of acceleration has been considered in the Minkowski space-time \cite{gedanken} as well as in the general relativistic regime \cite{grg,khomeriki}.

The magnetocentrifugal acceleration has been studied for the magnetospheres of  pulsars (please see the review paper \cite{osmanovRev} and references therein)  and black holes \citep{nikuradze25,osmanov07,rieger}.  These works examine a wide range of teh so-called astrophysical black holes and demonstrate the high efficiency of the acceleration process. However, prior studies have never considered primordial black holes with masses below the stellar-mass range \cite{hawking}. 

In this paper we study the acceleration of an electron in a black hole magnetosphere by focusing on black holes with masses in the range $(0.1-1)\times M_{\odot}$ , where $M_{\odot}\simeq 2\times 10^{33}$ g is the solar mass. 

The paper is organized  as follows: in Sec.2, we describe the mathematical model of particle acceleration, in Sec. 3 we apply our model to the mentioned class of primordial black holes to obtain results, and in Sec. 4, we summarize them. 

\section{Electron Dynamics}
 As already noted, due to the strong magnetic field, particles move along the magnetic field lines, which, to the first approximation, we treat as straight. Then, restricting the electron to move along the field line inclined by a constant angle $\theta_0$ with respect to the rotation axis, the Schwarzschild metric \cite{shapiro83}
\begin{equation}
ds^{2} = \left(1 - \frac{r_s}{r}\right) c^{2} dt^{2} - \frac{dr^{2}}{1 - \frac{r_s}{r}} - r^{2}(d\theta^{2} + \sin^{2}\theta \, d\varphi^{2}),
\end{equation}
will reduce to
\begin{equation}
ds^{2} = \left(1 - \frac{r_s}{r} - \frac{\omega^{2} r^{2} \sin^{2}\theta_0}{c^{2}}\right) c^{2} dt^{2} - \frac{dr^{2}}{1 - \frac{r_s}{r}},
\end{equation}
after the substitution $d\varphi = \omega dt$, $d\theta = 0$. Here $r_s$ denotes the Schwarzschild radius and $\omega$ is the angular velocity of rotation.

The fact that neither $\omega$ nor $\theta_0$ participate in this expression we define the effective angular velocity $\Omega=\omega\sin\theta_0$ . From the metric expression, one can directly read off the tensor components, which in matrix form are:

\begin{equation}
g_{\mu\nu} =
\begin{pmatrix}
1 - \frac{r_s}{r} - \frac{\Omega^{2} r^{2} }{c^{2}} & 0 \\
0 & -\dfrac{1}{1 - \frac{r_s}{r}}
\end{pmatrix}.
\end{equation}

An electron’s worldline is parametrized by a proper time in its rest frame $\tau= s/c$. Henceforth, overdots denote derivatives with respect to this parameter. In this parametrization, the action takes the form:
\begin{equation}
S = \int \left[ \left( 1 - \frac{r_s}{r} - \frac{\Omega^2 r^2}{c^2} \right)c^2 \dot{t}^2 
- \frac{\dot{r}^2}{1 - \frac{r_s}{r}} \right] d\tau
\end{equation}
leading to the following equations of motion
\begin{equation}
\ddot{r} = \left( \frac{\Omega^2 r}{c^2} - \frac{r_s}{2r^2} \right)\left(1 - \frac{r_s}{r}\right) c^2\dot{t}^2
+ \frac{1}{2} \frac{r_s \dot{r}^2}{r^2 - r_s r}
\end{equation}
\begin{equation}
\dot{t} = \frac{E}{1 - \frac{r_s}{r} - \frac{\Omega^2 r^2}{c^2}}.
\end{equation}
Here $E$ is an integration constant (the value of the conserved quantity associated with the t invariance). These equations can be reduced to the equation of motion for the flat space-time previously obtained by \citep{gedanken}:
\begin{equation}
\frac{d^2r}{dt^2}=\frac{\Omega^2r}{c^2-\Omega^2r^2}\left(c^2-\Omega^2r^2-2\left(\frac{dr}{dt}\right)^2\right).
\end{equation}
From this expression it is clear that in a nearby zone of the light cylinder $r\sim c/\Omega$, the acceleration becomes extremely efficient. The similar behavior takes place in a general relativistic treatment.

\section{Discussions and results}
Though its unstable, the equations of motion do admit one equilibrium position. In order to find it, we set both the radial acceleration and radial velocity to zero, i.e. $\ddot{r} = 0$ and $\dot{r} = 0$.  
This simplifies the equation to a balance between the gravitational attraction and the centrifugal term.  
Solving for $r$ gives us the radial position of this unstable equilibrium:

\begin{equation}
r_{\text{cr}} = \left(\frac{r_sc^2}{2\Omega^{2}}\right)^{1/3}.
\end{equation}

The unstable nature of the motion is evident from the form of the radial acceleration. It is straightforward to show that for $r>r_{cr}$ the centrifugal force dominates over gravity, causing the electron to be driven outward, whereas for  $r<r_{cr}$ the situation is reversed and the particle accelerates inward toward the black hole.

Another interesting behavior is hidden in the lab time $t$ derivative expression. There always exist zero or two positive roots of the denominator function in its expression. Which of those cases holds true depends on the sign of the cubic equation's discriminant, which gives rise to two cases: 1) $r_s>\frac{2c}{3\sqrt{3}\Omega}$ means that the expression of $\dot{t}$ always yields a negative number, which with the given metric results in a space-like interval; thus, a real physical particle cannot obey it. 2) $r_s>\frac{2c}{3\sqrt{3}\Omega}$ means that the derivative $\dot{t}$ is positive between the two positive roots of the mentioned equation, therefore, the metric is eligible to describe the electron motion.

Those roots can be written in the following form:
\begin{equation}
r_{\text{out}} = \frac{2c}{\sqrt{3}\,\Omega} \cos\!\left(\tfrac{1}{3} 
\arccos\!\left(-\frac{3\sqrt{3}\,r_s \Omega}{2c}\right)\right)
\end{equation}
\begin{equation}
r_{\text{in}} = \frac{2c}{\sqrt{3}\,\Omega} \cos\!\left(\tfrac{1}{3} 
\arccos\!\left(-\frac{3\sqrt{3}\,r_s \Omega}{2c}\right) - \tfrac{2\pi}{3}\right)
\end{equation}
Naming one inner and the other outer is a result of one always being larger than the other. It is worth noting that $r_{out}$ is a radius of a modified light cylinder (MLC). The concept of MLC has never been discussed in the literature. As expected, they coincide with the singularities predicted by the behavior of the Ricci scalar curvature. The unstable equilibrium position found before lies between them: $r_{in}<r_{cr}<r_{out}$. As the electron moves, it can approach both $r_{in}$ and $r_{out}$.In its proper frame, nothing prevents it from crossing these boundaries; however, as it approaches these singular radii, the proper time interval corresponding to a finite laboratory-frame time interval decreases without bound. As a result, infinitely long lab frame observations can only witness the behavior of the electron that happened before it crossed the singularity boundaries in its proper frame. Thus, an outside observer will see an electron asymptotically approaching the inner and outer radii, but never crossing them. Finally, all of this intuition can be backed up by the numerical solutions of the system of differential equations. 
\begin{figure}[H]
    \centering
    \includegraphics[width=1\linewidth]{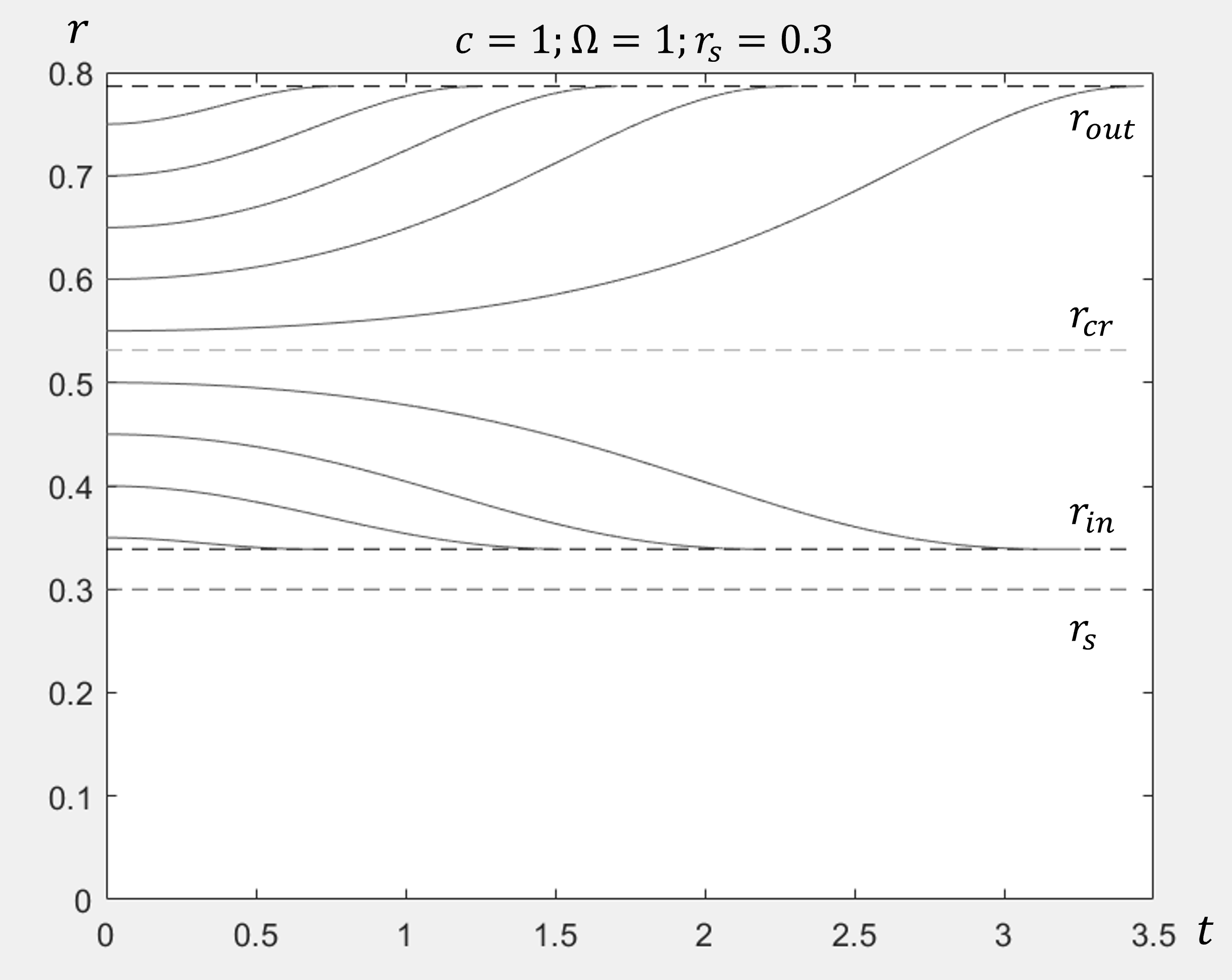}
    \ \caption{: Distance r from the center of a black hole as a function of lab time t.}
   
    \label{fig:enter-label}
\end{figure}

On Fig 1 nine different curves correspond to different initial positions ($r_0=0.35;0.40;0.45...0.75$) from where radially static electrons are released.As expected, the electrons approach one of the singularities depending on their initial positions. Electrons created between $r_c$ and $r_{in}$ asymptotically approach $r_{in}$while those created between $r_c$ and $r_{out}$  approach $r_{out}$  .It is worth noting that the boundary radii are functions of the angle $\theta_0$ of the field line's inclination. Knowledge of this angular dependence allows us to determine the shape of the region in which the metric remains timelike and can therefore describe the motion of an electron.

\begin{figure}[H]
    \centering
    \includegraphics[width=1\linewidth]{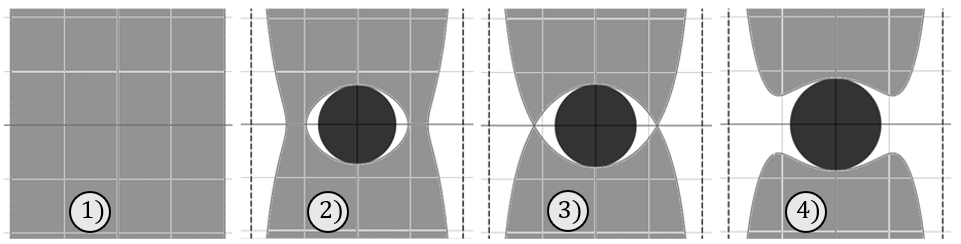}
    \ \caption{:Ergo sphere for different cases  1)$r_s = 0$  2)$r_s<\frac{2c}{3\sqrt{3}\Omega}$  3)$r_s=\frac{2c}{3\sqrt{3}\Omega}$ 4)$r_s>\frac{2c}{3\sqrt{3}\Omega}$ 
    }
   
    \label{fig:enter-label}
\end{figure}
It is instructive to examine how the “ergosphere”-like region evolves, as illustrated in Fig. 2. In the first case, a simple light cylinder is present. In the second, the light cylinder contracts and a central space-like region emerges. In the third case, the “ergosphere” begins to split at the equator. Finally, a finite separation is fully established.
 
 After a thorough analysis of the electron’s motion, we now turn to the energetic perspective. Defining the Lorentz factor as the time component of the four-velocity, we obtain the following expression:
\begin{equation}
\gamma = 
\frac{1}{\sqrt{1 - \tfrac{r_s}{r} - \left(\tfrac{\Omega r}{c}\right)^2 
- \tfrac{1}{c^2}\left(\frac{dr}{dt}\right)^2 \left(1 - \tfrac{r_s}{r}\right)}}
\end{equation}
Even though the equations of motion do not admit analytical expressions for nontrivial solutions, the symmetry associated with $\tau$ invariance makes it easy to find the dependence between the Lorentz factor and the distance $r$ during the electron's motion. Assuming the electron was released from $r_0$ with zero initial radial velocity the dependence assumes the following form:
\begin{equation}
\gamma = \frac{\sqrt{1 - \tfrac{r_s}{r_0} - \left(\tfrac{\Omega r_0}{c}\right)^2}}
{1 - \tfrac{r_s}{r} - \left(\tfrac{\Omega r}{c}\right)^2}
\end{equation}
The expression shows that the electrons' energies increase infinitely when they approach the boundaries of the "ergosphere". It coincides with the flat metric result by \cite{gedanken} at $r_s=0$.
\begin{figure}[H]
    \centering
    \includegraphics[width=1\linewidth]{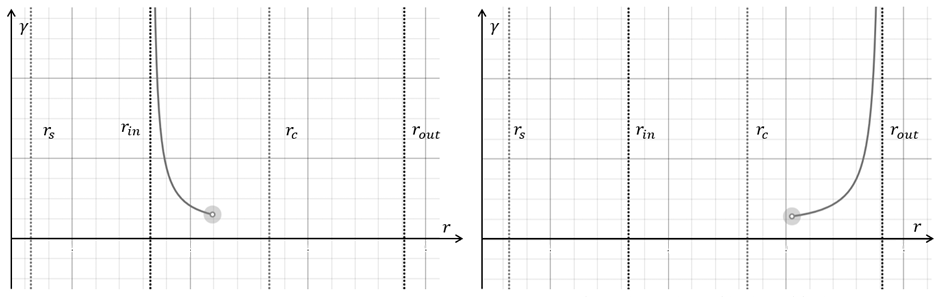}
    \ \caption{: Lorentz factor vs distance r for electrons released from both inside and outside the critical radius.}
   
    \label{fig:enter-label}
\end{figure}

{As we can see from Fig. 3 the expressions of Lorenz factors perfectly fit with results we got from Fig. 1 and we can clearly see that electrons that were in the inner region will achieve their maximum Lorenz factor when they approach $r_{in}$ and ones in the outer region $r_{out}$  }The model lacks any energy limiting factors that would cause dissipative terms in the equations of motion. The next part of this work will be dedicated to the upper limits of the electron's energies.
The limiting factors we are going to consider are the  inverse Compton scattering, curvature radiation and synchrotron emission. Taking these into account in our equation of motion would be an overly complex task. We already approximated the electron to move along a straight magnetic field line, supposedly giving up on the level of accuracy that would require us to investigate the exact dissipative forces. Thus one can be satisfied with a simplified energetical study.
Before considering the energy losses, we shall calculate the rate at which it increases. Since the equation of motion allows for representing the Lorentz factor $\gamma$ as a function of the electron's current distance from the black hole center and its initial coordinate, we are also able to calculate its growth rate with respect to the lab time $t$:
\begin{equation}
\frac{d\gamma}{dt}=\left| c\gamma\left(\frac{2\Omega^2r}{c^2}-\frac{r_s}{r^2}\right)\sqrt{\gamma_0\left(\gamma-\gamma_0\right)\left(1-\frac{r_s}{r}\right)}\right|
\end{equation}
Where $\gamma_0$ is the initial Lorentz factor of the released electron:
\begin{equation}
\gamma_0=\frac{1}{\sqrt{1-\frac{r_s}{r_0}-\left(\frac{\Omega r_0}{c}\right)^2}}
\end{equation}
Let us begin with the factor we concluded to be the dominant one for the mentioned mass range - the curvature radiation. It becomes important for extremely high Lorenz factors, that are achieved when the electron approaches its asymptotic coordinates (inner and outer radii). The curvature radiation power of an electron almost rotating on a circle is given by \cite{osmanovRev}:
\begin{equation}
P_{\text{cur}} = \frac{e^2c}{\rho_{cur}^2}\gamma^4,
\end{equation}
here $\rho_{cur}$ denotes the radius of curvature of the particle trajectory, which is of the order of the radius of MLC. At early times, the energy gain rate, $\dot{\gamma}mc^2$, exceeds the radiative loss rate. As the Lorentz factor increases, however, the power emitted via curvature radiation also grows. Consequently, the maximum attainable Lorentz factor is reached when the characteristic timescales of energy gain and energy loss become equal. Imposing this condition yields the following simple expressions on both boundaries:
\begin{equation}
\gamma_{out} \approx 
\left(
\frac{6\pi m_e}{e^2}
\left(
\frac{2\Omega^2}{c^2} r_{\text{out}}^3 - r_s
\right)
\sqrt{1 - \frac{r_s}{r_{\text{out}}}}
\right)^\frac{2}{5}\gamma_0^5\end{equation}

\begin{equation}
\gamma_{in} \approx 
\left(
\frac{6\pi m_e}{e^2 }
\left(r_s-
\frac{2\Omega^2}{c^2} r_{\text{in}}^3
\right)
\sqrt{1 - \frac{r_s}{r_{\text{in}}}}
\right)^\frac{2}{5}\gamma_0^5
\end{equation}
Another factor, which might influence the maximum attainable energies is the synchrotron radiation. The power emitted is given by \( p = \frac{2e^2 \omega_B^2 \gamma^2}{3c} \) \cite{rybicki}, where \( \omega_B = \frac{eB}{m_e c} \). From this, one can estimate the characteristic time for this process.
\begin{equation}
t_{\text{cool}} = \frac{\gamma mc^2}{p} = \frac{3m_e^3 c^5}{2 \gamma e^4 B^2}
\end{equation}
The magnetic field strength is determined under the assumption that the magnetic energy density is in equipartition with the radiative energy density of the accretion disk, $B \sim \sqrt{\frac{L}{ r^{2} c}}$ where luminosity is calculated by \cite{shapiro83}: 
\begin{equation}
L\simeq 1.2 \times 10^{30}\left(\frac{n_\infty}{1~\mathrm{cm^{-3}}}\right)^2\left( \frac{T_\infty}{10~\mathrm{K}} \right)^{-3}\left( \frac{M}{M_\odot} \right)^{3} \; [\mathrm{erg~s^{-1}}].
\end{equation}
Here the temperature and number density of particles are normalized by their values in the interstellar medium \( n_\infty \approx 1 \,\mathrm{cm}^{-3} \)
and \( T_\infty \approx 10\,\mathrm{K} \). By equating the time-scales, one arrives at:
\begin{equation}
\gamma^{\mathrm{in}}_{\mathrm{syn}}
=
\frac{
9\, m_e^{6}\, c^{14}
\left(
\frac{2 \Omega^{2} r_{\mathrm{in}}^{3}}{c^{2}} - r_s
\right)^{2}
}{
4\, e^{8}
\left(
1.2 \times 10^{30}
\left(
\frac{M}{M_\odot}
\right)^{3}
\right)^{2}
}
\end{equation}
and 
\begin{equation}
\gamma^{out}_{syn}
= \frac{9 m_e^6 c^{14} \left( \frac{2 \Omega^2 r_{out}^3}{c^2} - r_s \right)^2}
           {4 e^8 \left( 1.2 \times 10^{30} \left( \frac{M}{M_\odot} \right)^3 \right)^2 }
\end{equation}

We now turn to estimating the exact values of Lorentz factors attainable by electrons in the vicinity of black holes. 
For this analysis, as we have already mentioned, we focus on \textit{primordial black holes} due to their distinctive physical properties, 
particularly their comparatively low masses. In particular, we consider a mass range extending from one solar mass 
(\(M_\odot\)) down to one-tenth of the solar mass (\(0.1\,M_\odot\)).
We can now calculate values that were limited by curvature radiation:
\begin{figure}[H]
    \centering
    \includegraphics[width=1\linewidth]{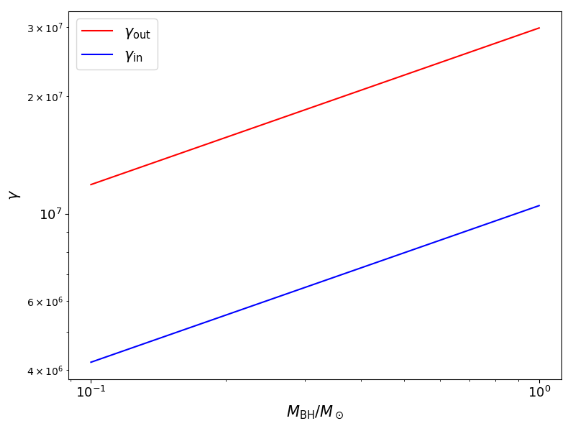}
    \ \caption{: Lorentz Factors(Curvature radiation) vs Black Hole Mass}
   
    \label{fig:enter-label}
\end{figure}
As can be seen, particles created in the inner region attain significantly lower energies than those formed in the outer layer. This behavior is consistent with the analytical expression and is therefore not unexpected. We also clearly observe that the particle energy increases with black hole mass, supporting the notion that curvature radiation becomes a less effective limiting mechanism in supermassive black holes.
From formula (20) and (21) we can numerically calculate and plot, maximum Lorenz factor with this regime.  
\begin{figure}[H]
    \centering
    \includegraphics[width=1\linewidth]{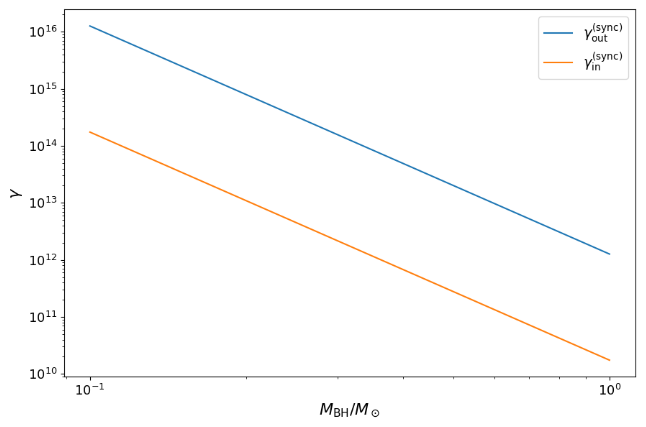}
    \ \caption{: Lorentz Factors(synchrotron radiation ) vs Black Hole Mass}
   
    \label{fig:enter-label}
\end{figure}
From the graph it is evident that the maximum Lorentz factors limited by synchrotron radiation are much higher than those limited by curvature radiation. 
{
Now we consider the inverse Compton scattering. In particular, we should examine the regimes of the process. To do this, we need to estimate the quantity \( \frac{\gamma \epsilon_{\text{ph}}}{m_e c^2} \), where \( \epsilon_{\text{ph}} \) is the photon's energy. If this quantity is much smaller  than unity, the scattering is in the Thomson regime, and if it is much greater than unity, the scattering is in the Klein-Nishina regime. To estimate the photon energies, we can apply Wien's law for black body radiation.

\begin{equation}
\lambda_{\text{peak}} T \simeq 0.3  \, \text{cm}\; \text{K}
\end{equation}

Here, \( T \) represents the temperature of the photon gas, and \( \lambda_{\text{peak}} \) denotes the wavelength of the photons. For a typical temperature of the accretion disk, one has \cite{shapiro83}:
\begin{equation}
T \approx 1.2 \times 10^{7}
\left( \frac{10\,M_\odot}{M} \right)^{-1/4}
\left( \frac{3 r_s}{r} \right)^{-3/4}
\ \mathrm{K}.
\end{equation}
As it turns out, a quantity \(\frac{\gamma \epsilon_{\text{ph}}}{m_ec^2} = \frac{h\,T}{b\,m_e\,c}\), is of the order of $10^8-10^9$, implying that the Klein-Nishina regime is efficient, and the corresponding power is given by \cite{blumenthal}:
\begin{equation}
\
 P_{KN} = \frac{\sigma_T {\left( m_eckT \right)}^2}{16\hbar^3}\left( \ln{\left( \frac{4\gamma k T}{m c^2} \right)} -1.981\right){\left( \frac{r_s}{r} \right)}^2
\end{equation}
and determine the maximum Lorentz factors, we employ the same technique as before. Using this expression, we can then estimate the characteristic cooling time near the MLC associated with this limiting factor\[
t_{\mathrm{IC}} = \frac{\gamma\, m_e c^2}{P_{c,\mathrm{KN}}} \propto \gamma
\]
which, to first order, is proportional to $\gamma$. Acceleration, on the
other hand, occurs on a timescale $t_{\mathrm{acc}} \propto \gamma^{-1/2}$ \citep{osmanovRev},
so that for electron Lorentz factors $\gamma \gtrsim \gamma_0$, the inverse Compton cooling does not impose any constraint on the achievable particle energies.
Therefore, it is reasonable to conclude that curvature radiation serves as the primary constraint on electron acceleration.
}

\section{Conclusion}
In this work, we extended the classical concept of the light cylinder by including gravitational effects on a charged particle co-rotating along an inclined magnetic field line. Then, in the Schwarzschild metric we derived an effective two-dimensional metric that reveals \textit{two light-cylinder--like radii}, $r_{\rm in}$ and $r_{\rm out}$, separated by an \textit{unstable critical orbit}, $r_{\rm cr}$. The region between these radii behaves like an "ergosphere": particles originating inside the inner or outer regions tend to asymptotically approach $r_{\rm in}$ or $r_{\rm out}$ as seen by a distant observer and cannot cross these surfaces in finite lab time.  

We obtained analytic expressions for both the critical radius and the light-cylinder radii and used the conserved quantities of the effective metric to track the evolution of particle Lorentz factors along their trajectories. Focusing on low-mass (primordial) black holes in the range $0.1\,M_\odot$ to $1\,M_\odot$, we found that \textit{curvature radiation dominates the limitation on electron acceleration}, with synchrotron losses remaining subdominant over this mass range.  

Overall, the present model generalizes the light-cylinder concept to a \textit{relativistic gravitational setting}, providing a simple framework to study particle acceleration and plasma behavior around compact objects where both gravity and rotation are important. Future work could extend this approach to include \textit{frame-dragging effects} (Kerr spacetimes), \textit{self-consistent electromagnetic fields}, or \textit{kinetic/MHD effects} to explore collective plasma dynamics within this generalized "ergosphere".

\section*{Acknowledgements}

This work was supported by the Shota Rustaveli National Science Foundation of Georgia under Grant No. FR-24-1751.

\section*{Data availability}
Data are available in the article and can be accessed via a DOI link.

\
\end{document}